\documentclass[12pt]{article}
\usepackage{times}
\usepackage{epsf}
\usepackage{graphicx}

\def\beq{\begin{equation}}
\def\eeq{\end{equation}}
\def\bea{\begin{eqnarray}}
\def\eea{\end{eqnarray}}
\def\ba{\begin{array}}
\def\ea{\end{array}}
\def\bea{\begin{eqnarray}}
\def\eea{\end{eqnarray}}

\begin{document}
\title{{\bf Exact Polynomial Solutions of the Mie-Type Potential in the
N--Dimensional Schr\"{o}dinger Equation }}
\author{Sameer Ikhdair \thanks{sikhdair@neu.edu.tr },
Ramazan Sever \thanks{sever@metu.edu.tr } \\[1cm]
$^{*}$Department of Physics, \ Near East University, Nicosia, North\\
Cyprus, Mersin-10, Turkey  \\[.5cm]
$^ \dagger$Department of Physics, Middle East\\
Technical \ University, 06531 Ankara, Turkey}

\date{\today}
\maketitle \normalsize

\begin{abstract}
The polynomial solution of the $N$-dimensional space Schr\"{o}dinger
equation for a special case of Mie potential is obtained for any arbitrary $%
l $-state. The exact bound-state energy eigenvalues and the
corresponding eigenfunctions are calculated for diatomic molecular
systems in the Mie-type potential. Keywords: Mie potential,
Schr\"{o}dinger equation, Eigenvalue, Eigenfunction, Diatomic
molecules PACS\ number: 03.65.-w; 03.65.Fd; 03.65.Ge.
\end{abstract}
\newpage
\section{Introduction}
\noindent The solution of the Schr\"{o}dinger equation for any
spherically symmetric potential has attracted attention over the
past years [1-16]. The motivation in this direction arises from
considerable applications in the different fields of the material
science and solid physics.

The Harmonic oscillator and H-atom (Coulombic) problems are two
exactly solvable potentials which have been thoroughly studied in
$N$-dimensional space quantum mechanics for any angular momentum
$\ell$. These two problems are related together and hence the
resulting second-order differential equation has the normalized
orthogonal polynomial function solution (cf. Ref.[17] and the
references therein). On the other hand, the Pseudoharmonic and Mie
type potentials are also two exactly solvable potentials other than
Coulombic and Harmonic oscillator, their wavefunctions are vanishing
at the origin.

The path integral solution for one-dimensional special case Mie-potential
which is a perturbed Coulombic-type potential was obtained in [18].
Moreover, the Schr\"{o}dinger equation for a system bound by a Mie-type
potential was also solved by using the $1/N$ expansion method [19].

In this letter we will follow parallel solution to Refs.[17, 20] and
give a complete normalized polynomial solution for the general
$N$-dimensional Schr\"{o}dinger equation for diatomic molecular
systems interacting through Mie type potential, Coulombic-type
potential with an additional centrifugal potential barrier, which
reduces to the standard three dimensional case when the parameter
$N$ is set equal to $3.$ Our aim is to present a different
approach to calculate the non-zero angular momentum solutions of the $N$%
-dimensional space of Schr\"{o}dinger equation for the Mie-type potential.

The contents of this paper is as follows. In Section \ref{TND}, we give the
eigensolution of the $N$ dimensional Schr\"{o}dinger equation with Mie type
potential for the analytical bound-state (real) and imaginary solutions.
Finally, in Section \ref{RAC} we give our results and conclusions.

\section{The $N$-Dimensional Schr\"{o}dinger Equation with Mie-Type Potential%
}

\label{TND}We wish to solve the $N$-dimensional Schr\"{o}dinger equation for
a special case of the Mie-type potential [18] given by

\begin{equation}
V(r)=D_{0}\left[ \frac{k}{m-k}\left( \frac{r_{0}}{r}\right) ^{m}-\frac{m}{m-k%
}\left( \frac{r_{0}}{r}\right) ^{k}\right] ,
\end{equation}
where $D_{0}$ is the interaction energy between two atoms in a solid at $%
r=r_{0},$ and $m>k$ is always satisfied. By taking $m=2k$ and choosing the
special case $k=1,$ corresponding to a Coulombic-type potential with
additional centrifugal potential barrier, Eq.(1) becomes

\begin{equation}
V(r)=-\frac{a_{1}}{r}+\frac{a_{2}}{2r^{2}},
\end{equation}
\begin{equation}
r^{2}=\sum_{i=1}^{N}x_{i}^{2},
\end{equation}
where $a_{1}=2D_{0}r_{0}$ and $a_{2}=2D_{0}r_{0}{}^{2}.$ For brevity, we
write the radial part of the Schr\"{o}dinger equation in $N$-dimensions as

\begin{equation}
\left[ -\frac{\hbar ^{2}}{2m}{\bf \nabla }_{N}^{2}+V(r)\right] \psi
(r)=E_{nl}\psi (r),
\end{equation}
which reduces into the form

\begin{equation}
\left\{ \frac{d^{2}}{dr^{2}}+\frac{N-1}{r}\frac{d}{dr}-\frac{l(l+N-2)}{r^{2}}%
+\frac{2m}{\hbar ^{2}}\left[ E_{nl}+\frac{a_{1}}{r}-\frac{a_{2}}{2r^{2}}%
\right] \right\} R_{nl}(r)=0.
\end{equation}
Using the dimensionless abbreviations:

\begin{equation}
x=r/r_{0};\beta ^{2}=-\frac{2mr_{0}^{2}}{\hbar ^{2}}E_{nl};%
\gamma ^{2}=\frac{2mr_{0}^{2}}{\hbar ^{2}}D_{0},
\end{equation}
gives the following simple form

\begin{equation}
\frac{d^{2}R_{nl}(x)}{dx^{2}}+\frac{N-1}{x}\frac{dR_{nl}(x)}{dx}+\left[
-\beta ^{2}+\frac{2\gamma ^{2}}{x}-\frac{\gamma ^{2}+l(l+N-2)}{x^{2}}\right]
R_{nl}(x)=0,
\end{equation}
which restores its three-dimensional Schr\"{o}dinger standard form once $N=3$%
. Eq. (7) has an irregular singularity in the $x\rightarrow \infty $
limit where its normalizable solutions in bound-states behave as
[17, 20]
\begin{equation}
\left( \frac{d^{2}}{dx^{2}}-\beta ^{2}\right) R_{nl}(x)=0,
\end{equation}
so that $R_{nl}(x)=A_{nl}\exp (-\beta x)+B_{nl}\exp (\beta x)$ $\ $In order
that $R_{nl}(x)\rightarrow 0$ as $x\rightarrow \infty ,$ we set $B_{nl}=0,$
so

\begin{equation}
\mathrel{\mathop{\lim }\limits_{x\rightarrow \infty }}%
R_{nl}(x)=A_{nl}\exp (-\beta x),
\end{equation}
where $A_{nl}$ is the normalization constant. Eq. (8) leads us to
propose the trial solution

\begin{equation}
R_{nl}(x)=N_{nl}\exp (-\beta x)g(x),
\end{equation}
where $N_{nl}$ is another normalization constant. Putting this back into Eq.
(7) yields an equation for $g(x)$ of the form
\begin{equation}
{g}^{\prime \prime }(x)+\left( \frac{N-1}{x}-2\beta \right)
g^{\prime
}(x)+\left( \frac{2\gamma ^{2}-(N-1)\beta }{x}-\frac{\gamma ^{2}+l(l+N-2)}{%
x^{2}}\right) g(x)=0,
\end{equation}
where the prime refers to the derivative with respect to $x.$ Because
exponential behavior has already been taken out, one hopes that the solution
for $g(x)$ is a polynomial. Indeed, Eq. (7) has a singularity at $%
x\rightarrow 0,$ the substitution of the trial solution $g(x)=x^{q},$
provides the positive root solution:

\begin{equation}
q=-\frac{(N-2)}{2}+\sqrt{\left( l+\frac{N-2}{2}\right) ^{2}+\gamma ^{2}}.
\end{equation}
As $q>0,$ the wavefunction vanishes at $x=0,$ corresponding to the strong
repulsion between the two atoms. It is reasonable to set

\begin{equation}
R_{nl}(x)=N_{nl}x^{q}\exp (-\beta x)h(x),
\end{equation}
to Eq. (7), one gets

\[
h^{\prime \prime }(x)+\left( \frac{2q+N-1}{x}-2\beta \right) h^{\prime }(x)
\]
\begin{equation}
+\left( \frac{2\gamma ^{2}-2q\beta -(N-1)\beta }{x}+\frac{%
q(q-1)+(N-1)q-\gamma ^{2}-l(l+N-2)}{x^{2}}\right) h(x)=0.
\end{equation}
Setting the numerator of $x^{-2}$ term equal to zero$,$ in the last
equation, and solving the resulting quadratic equation leads again
to the solution given in Eq. (12) and thus gives the following
differential equation

\begin{equation}
xh^{\prime \prime }(x)+\left[ 2q+N-1-2\beta x\right] h^{\prime }(x)+\left[
2\gamma ^{2}-2q\beta -(N-1)\beta \right] h(x)=0.
\end{equation}
The confluent series, for large values of $x,$ is proportional to $\exp
(2\beta x)$ so that $R_{nl}(x)$ diverges for $x\rightarrow \infty $ if the
series $_{1}F_{1}$ does not break off. If it does $_{1}F_{1}$ is a
polynomial and $R_{nl}(x)\rightarrow 0$ for $x\rightarrow \infty $ becomes
normalizable. After the substitution of series form [20]
\begin{equation}
h(x)=\sum_{i=0}^{i_{\max }}C_{i}x^{i},
\end{equation}
into Eq. (15), it provides

\begin{equation}
C_{i+1}=\frac{i+q+(N-1)/2-\gamma ^{2}/\beta
}{(i+1)(i+2q+N-1)}C_{i}=\Gamma _{il}C_{i},
\frac{C_{i+1}}{C_{i}}\rightarrow \frac{1}{i},
\end{equation}
which leads to a divergent wavefunction if not truncated to a maximum value
for $i.$ Neverthless, the wavefunction is finite everywhere since $i$ and $l$
are finite, it follows

\begin{equation}
i_{\max }+q+\frac{(N-1)}{2}-\frac{\gamma ^{2}}{\beta }=0; i_{\max }=n;%
\left( n=0,1,2,...\right) .
\end{equation}
Further, using the following abbreviations in Eq. (15):

\begin{equation}
z=2\beta x;c=\left( q+\frac{N-1}{2}\right) ;%
-n=\left( c/2-\gamma ^{2}/\beta \right) ,
\end{equation}
gives the general type of Kummer's (Confluent Hypergeometric) differential
equation of the form

\begin{equation}
zh^{\prime \prime }(z)+\left[ c-z\right] h^{\prime }(z)+nh(x)=0;%
\left( n=0,1,2,...\right) ,
\end{equation}
having the solution

\begin{equation}
h(x)=_{1}F_{1}(a,c;z)=_{1}F_{1}(-n;2\frac{\gamma ^{2}}{\beta }-2n;2\beta
x)=L_{n}^{(2\frac{\gamma ^{2}}{\beta }-2n-1)}(2\beta x),
\end{equation}
where $_{1}F_{1}(-n,m+1;z)=L_{n}^{(m)}(z)$ denotes the Kummer's
function. On the other hand, we may also write Eq. (15) in the
following general form

\begin{equation}
zh^{\prime \prime }(z)+\left[ m+1-z\right] h^{\prime }(z)+nh(x)=0,
\end{equation}
where $m=2q+N-2$ and $n_{\max }=\frac{\gamma ^{2}}{\beta }-q-\frac{(N-1)}{2}%
=0,1,2,....$ If $n$ is a non-negative integer, then a finite
polynomial solution is allowed. (This when combined with the rest of
$R_{nl}(x)$ yields a normalizable solution.) In particular, this
solution to Eq. (22) is the generalized Laguerre polynomial
$L_{n}^{(m)}(2\beta x).$ Combining everything one finally has

\begin{equation}
R_{nl}(r)=N_{nl}\left( \frac{r}{r_{0}}\right) ^{\frac{\gamma ^{2}}{\beta }-n-%
\frac{(N-1)}{2}}\exp \left( -\frac{\beta r}{r_{0}}\right) L_{n}^{(2\frac{%
\gamma ^{2}}{\beta }-2n-1)}\left( \frac{2\beta r}{r_{0}}\right) .
\end{equation}
Since the radial volume element in $N$-dimensional space is $r^{N-1}dr,$ one
can obtain [17]

\[
N_{nl}=\left[ \int_{0}^{\infty }drr^{N-1}e^{-2\beta x}x^{2q}\left(
L_{n}^{(2q+N-2)}(2\beta x)\right) ^{2}\right] ^{-1/2},
\]

\begin{equation}
N_{nl}=\frac{\left( 2\beta \right) ^{q+N/2}}{r_{0}^{N/2}}\left[
J_{n,m}^{(1)}(2\beta x)\right] ^{-1/2};J_{n,m}^{(1)}=\frac{(m+n)!}{n!%
}(2n+m+1),
\end{equation}
or equivalently as

\begin{equation}
N_{nl}=\left( \frac{2\beta }{r_{0}}\right) ^{N/2}\left( 2\beta \right) ^{%
\frac{\gamma ^{2}}{\beta }-n-\frac{(N-1)}{2}}\left[ \frac{n!}{\frac{2\gamma
^{2}}{\beta }\Gamma (\frac{2\gamma ^{2}}{\beta }-n)}\right] ^{1/2}.
\end{equation}

\subsection{Negative Energy}

For bound-states, $\beta >0,$ the solution becomes in arbitrary
normalization constant is given by means of Eq. (23) in which $R_{nl}(x)\rightarrow 0$ for $%
x\rightarrow \infty $ becomes normalizable. On the other hand, from
Eq. (6) with the aid of Eqs. (12) and (18), the eigenvalue becomes
[21]

\begin{equation}
E_{nl}=-\frac{\hbar ^{2}\gamma ^{4}}{2mr_{0}^{2}}\left[ n+\frac{1}{2}+\sqrt{%
\left( l+\frac{N-2}{2}\right) ^{2}+\gamma ^{2}}\right] ^{-2}.
\end{equation}
Since the parameter $\gamma \gg 1$ for most diatomic molecules, we may
expand into powers of $1/\gamma .$ This leads to

\begin{equation}
E_{nl}=D_{0}\left[ -1+\frac{2\left( n+\frac{1}{2}\right) }{\gamma }+\frac{%
\left( l+\frac{N-2}{2}\right) ^{2}}{\gamma ^{2}}-\frac{3\left( n+\frac{1}{2}%
\right) ^{2}}{\gamma ^{2}}-\frac{3\left( n+\frac{1}{2}\right) \left( l+\frac{%
N-2}{2}\right) ^{2}}{\gamma ^{3}}+...\right] .
\end{equation}
The Mie-type potential, Eq.(2), can be expanded about its minimum at $r=r_{0}
$ as
\begin{equation}
V(r)=D_{0}\frac{(r-r_{0})^{2}}{r_{0}^{2}}-D_{0}.
\end{equation}
Therefore, with classical frequency for small harmonic vibrations,

\begin{equation}
\omega =\sqrt{\frac{2D_{0}}{mr_{0}^{2}}},
\end{equation}
and the moment of inertia

\begin{equation}
I=mr_{0}^{2},
\end{equation}
we arrive at

\[
E_{nl}=-\frac{1}{2}I\omega ^{2}+\hbar \omega \left( n+\frac{1}{2}\right) +%
\frac{\hbar ^{2}}{2I}\left( l+\frac{N-2}{2}\right) ^{2}-\frac{3\hbar ^{2}}{2I%
}\left( n+\frac{1}{2}\right) ^{2}
\]

\begin{equation}
-\frac{3\hbar ^{3}}{2I^{2}\omega }\left( n+\frac{1}{2}\right) \left( l+\frac{%
N-2}{2}\right) ^{2}+....
\end{equation}

\subsection{Positive Energy}

For $\beta <0$ or $E_{nl}>0,$ it is no longer real but purely complex, $%
\beta x=-i\kappa r$ with $\kappa =\sqrt{\frac{2mE}{\hbar ^{3}}}$
which gives the wavefunction [21]

\begin{equation}
R_{nl}(r)=A_{nl}(r/r_{0})^{q}\exp (i\kappa r)_{1}F_{1}(q+\frac{N}{2}-\frac{1%
}{2}-\frac{i\gamma ^{2}}{\kappa r_{0}},2q+N-1;-2i\kappa r),
\end{equation}
which vanishes at $r=0.$ Its asymptotic may be found from the well-known
formula [21]

\begin{equation}
_{1}F_{1}(a,c;z)\rightarrow \exp (-i\pi a)\frac{\Gamma (c)}{\Gamma (c-a)}%
z^{-a}+\frac{\Gamma (c)}{\Gamma (a)}\exp (z)z^{a-\eta _{l}}=0,
\end{equation}
holding for the whole complex $z$-plane cut along the positive imaginary
axis.

\[
R_{nl}(r)=C_{nl}r^{q}\exp (i\kappa r)\left[ e^{-i\pi \left( q-\frac{i\gamma
^{2}}{\kappa r_{0}}+\frac{N}{2}-\frac{1}{2}\right) }\frac{\Gamma \left(
2q+N-1\right) }{\Gamma \left( q+\frac{i\gamma ^{2}}{\kappa r_{0}}+\frac{N-1}{%
2}\right) }(-2i\kappa r)^{-q+\frac{i\gamma ^{2}}{\kappa r_{0}}-\frac{(N-1)}{2%
}}\right.
\]
\begin{equation}
\left. +\frac{\Gamma \left( 2q+N-1\right) }{\Gamma \left( q-\frac{i\gamma
^{2}}{\kappa r_{0}}+\frac{N-1}{2}\right) }e^{-2i\kappa r}\left( -2i\kappa
r\right) ^{q-\frac{i\gamma ^{2}}{\kappa r_{0}}+\frac{N-1}{2}-\eta _{l}}%
\right] ,
\end{equation}
with $\frac{\gamma ^{2}}{\kappa r_{0}}=\sqrt{\frac{2m}{\hbar ^{3}}\frac{%
D_{0}^{2}}{E}}r_{0}.$ Therefore, the eigenenergy reads as
\begin{equation}
E_{nl}=\frac{\hbar ^{2}\gamma ^{4}}{2mr_{0}^{2}}\left[ n+\frac{1}{2}+\sqrt{%
\left( l+\frac{N-2}{2}\right) ^{2}+\gamma ^{2}}\right] ^{-2}.
\end{equation}

\section{Results and Conclusions}

\label{RAC}We have studied the solution for a Mie-type potential.
Considering the special case of the Mie potential with $m=2k$ and
for $k=1,$ the problem was reduced to a Coulombic potential with the
additional centrifugal potential barrier of order $1/r^{2}.$ The
exact solutions for this particular case have been obtained, which
are similar to the Hydrogenic solutions [17]. We have calculated the
eigenvalue and the corresponding wave function considering the
negative (bound-states) and positive (imaginary) cases for any
quantum-mechanical system bound by a special case of the Mie
potential.

The present results for the potential parameters $a_{1}=1$ and
$a_{2}=0$ in Eq. (2) correspond to the Coulombic potential case.
They are the same with the previous calculations in [17-19]. Thus,
for this particular case the energy terms in the expansion, Eq.
(27), take the form

\begin{equation}
E_{n}=-\frac{1}{n^{2}}.
\end{equation}
This value is the exact nonrelativistic H-atom energy expression. Further,
setting $N=3,$ $D_{0}=V_{0}/2,$ $r_{0}=\sigma ,$ $A=\frac{1}{2}\sigma
^{2}V_{0}$ and $B=\sigma ^{2}V_{0},$ Eq.(27) becomes

\begin{equation}
E_{n}=-\frac{2m}{\hbar ^{2}}B^{2}\left[ 2n^{\prime }+1+\left[ \left(
2l+1\right) ^{2}+\frac{8mA}{\hbar ^{2}}\right] ^{1/2}\right] ^{-2},%
n^{\prime }=n-s-1=1,2,3...
\end{equation}
which consequently recovers the formula (28) in Ref. [18] and also
Eq. (19) in Ref. [19]. For the H-atom case this energy expression
also gives $-0.5$ a.u. exactly. Thus, the present result reproduces
exactly the path integral, $1/
N$%
-expansion, and the non-relativistic Schr\"{o}dinger equation
solutions.

\section{Acknowledgments} This research was partially supported by the
Scientific and Technological Research Council of Turkey. S.M.
Ikhdair wishes to dedicate this work to his family for their love
and assistance.

\end{document}